\documentclass[aip,apl,groupedaddress,amsmath,amssymb,reprint]{revtex4-1}
\usepackage{graphicx}
\usepackage{bm}

\begin{document}


\title{High mobility two-dimensional hole system on hydrogen-terminated silicon (111) surfaces} 



\author{Binhui Hu}
\email[To whom correspondence should be addressed. Electronic mail:
] {hubh@lps.umd.edu}
\author{Tomasz M. Kott}
\author{Robert N. McFarland}
\altaffiliation [Present address: ] {i\_SW, Arlington, VA 22203,
USA.}
\author{B. E. Kane}
\affiliation{Laboratory for Physical Sciences, University of
Maryland at College Park, College Park, MD 20740}
\date{\today}

\begin{abstract}
We have realized a two-dimensional hole system (2DHS), in which the
2DHS is induced at an atomically flat hydrogen-terminated Si(111)
surface by a negative gate voltage applied across a vacuum cavity.
Hole densities up to $7.5\times10^{11}$ cm$^{-2}$ are obtained, and
the peak hole mobility is about $10^4$ cm$^2$/Vs at 70 mK. The
quantum Hall effect is observed. Shubnikov-de Haas oscillations show
a beating pattern due to the spin-orbit effects, and the inferred
zero-field spin splitting can be tuned by the gate voltage.
\end{abstract}


\maketitle 


Metal-oxide-semiconductor (MOS) field-effect transistors (FETs) have
long been used to study two-dimensional systems in silicon. While
there has been extensive research on two-dimensional electron
systems (2DESs) in silicon using MOSFETs over the last several
decades,\cite{KlitzingQHE80, Ando2DS82} comparably little work has
been done on two-dimensional hole systems
(2DHSs),\cite{KlitzSi11074, KlitzSi11174, Ando2DS82, [{}][{ [JETP
Lett. 46, 502 (1987)]}] Dorozhkin87} primarily due to the fact that
holes have lower mobility than electrons, and detailed study was not
possible at such low mobility silicon hole devices.

Recently, a new type of silicon FET device has been developed, in
which an atomically flat silicon (111) surface is terminated by a
monolayer of hydrogen atoms, and a 2DES is induced at the H-Si(111)
surface by a positive gate voltage through a vacuum
barrier.\cite{Eng05} 2DESs in these vacuum FET devices show
extremely high mobility.\cite{Eng05,Eng07,Robert09} In this letter,
we report that the vacuum FET technique can be extended to 2DHSs. A
peak hole mobility of $\sim$ 10$^4$ cm$^2$/Vs is obtained, exceeding
by one order of magnitude the mobility in MOS structures and
comparable with the mobility of pseudomorphic Si/SiGe
heterostructures,\cite{whall94,*basaran94,[{Strained Ge channels on
SiGe virtual substrates can reach a higher hole mobility of $\sim$
120000 cm$^2$/Vs at 2 K, as shown in }][{}]rossner04} where 2DHSs
reside in the SiGe channels. The integer quantum Hall effect (IQHE)
is also observed. At lower magnetic fields, Shubnikov-de Haas (SdH)
oscillations show a beating pattern, which is due to the zero-field
spin splitting.\cite{Zawadzki04} In silicon, the spin-orbit effects
lift the spin-degeneracy of the heavy hole subband even in a zero
external magnetic field. The zero-field spin splitting is from the
structure inversion asymmetry (SIA),\cite{WinklerSO} and can be
tuned by the external gate. The zero-field spin splitting is of
great interest because of its potential use in spintronics
devices\cite{Datta90} and in studying fundamental
physics.\cite{WinklerSO}

\begin{figure}
\includegraphics{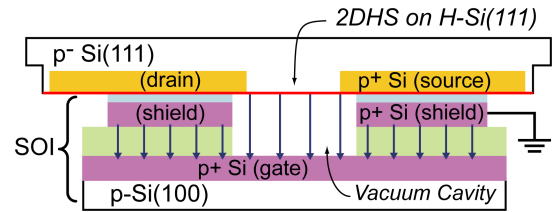}
\caption{\label{fig:schematics} (Color online) Schematic cross
section of a vacuum FET (not to scale). A H-Si(111) piece is contact
bonded to a SOI piece in a vacuum. The H-Si(111) piece has heavily
boron-doped (p$^+$) regions to make contact to the 2DHS. In the SOI
piece, the shield and the gate are p$^+$ layers formed by double
boron ion implantation. Dry etching is used to create the cavity.
The arrows depict the electric field. The 2DHS is induced at the
H-Si(111) surface in the encapsulated vacuum cavity.}
\end{figure}

\begin{figure}
\includegraphics{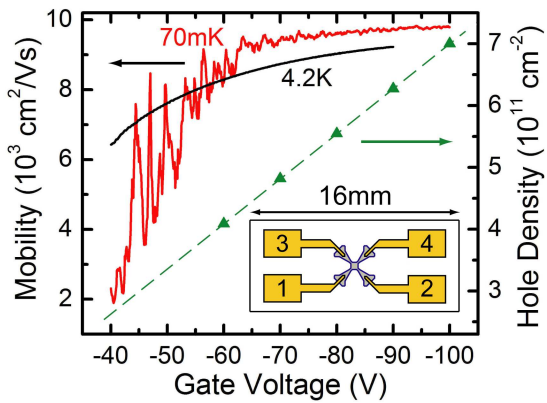}
\caption{\label{fig:mobdensity} (Color online) Left axis: mobility
as a function of gate voltage at 70 mK and 4.2 K. At 70 mK, the wild
fluctuations of the mobility are due to the high contact resistances
to the 2DHS. Right axis: Hole density vs gate voltage. The hole
densities (triangles) are determined by the magnetic field B at the
filling factor $\nu=5$ minima: $p_{2d}=\nu eB/h$. The dashed line is
a linear fit. Insert: on the Si(111) piece, Van der Pauw geometry
consists of a cloverleaf-shaped 2DHS with a $0.5\times0.5$ mm$^2$
center square, and four heavily boron-doped contacts.}
\end{figure}

The vacuum FET device includes two pieces, as shown in Fig.\
\ref{fig:schematics}. One is a p$^-$ Si(111) piece (float zone,
resistance $>10000\ \Omega\cdot$cm), and the other is a
silicon-on-insulator (SOI) piece. First, four p$^+$ contact regions
are formed in the Si(111) piece by ion implantation with
$2.4\times10^{15}$ cm$^{-2}$, 15 keV boron ions through a 30 nm
thick thermal oxide, as shown in the insert of Fig.\
\ref{fig:mobdensity}. It is annealed in N$_2$ gas at 1000$^\circ$C
for 10 minutes to activate dopants and reduce defects. Second, in
the SOI piece, the shield and gate conducting layers are formed by
double boron ion implantation, and a cloverleaf-shaped cavity is
created by dry etching. Finally, the Si(111) piece is H-terminated
by immersing it in an ammonium fluoride solution, and then these two
pieces are contact bonded in a vacuum chamber. The detailed
fabrication processes and the basic operating principle have been
discussed elsewhere.\cite{Eng05, RobertTh10} Here the SOI piece acts
as a remote gate to induce a 2DHS at the Si(111) surface through the
encapsulated vacuum cavity. This cavity also protects the air
sensitive H-Si(111) surface. Compared to the Si/SiO$_2$ interface in
MOSFETs, the vacuum/H-Si(111) interface is atomically flat and has
much fewer defects, leading to a much higher carrier mobility.

The devices are characterized in a dilution refrigerator with a base
temperature of $\sim$ 70 mK. Longitudinal ($R_{xx}$) and Hall
($R_{xy}$) resistances are determined by Van der Pauw measurements
with the standard low-frequency AC lock-in technique using a 7 Hz,
100 nA current source. As shown in the insert of Fig.\
\ref{fig:mobdensity}, the Van der Pauw geometry consists of a center
cloverleaf-shaped 2DHS, which is electric field induced, and four
heavily boron-doped contacts (labeled as 1, 2, 3, 4). Four-terminal
resistance is defined as $R_{ij,lm}=V_{lm}/I_{ij}\
(i,j,l,m=1,2,3,4)$, where current $I_{ij}$ is fed from contact i to
contact j, and voltage $V_{lm}$ is measured between contacts l and
m. The sheet resistance is determined by the standard Van der Pauw
technique.\cite{vdP58} Although the Van der Pauw geometry is
symmetric, $R_{xx}$ and $R_{yy}$ are found to differ by $\sim$ 35\%,
so the sheet resistance, as well as the mobility, is the average of
x and y directions.\cite{Bierwagen04} The anisotropy can be caused
by some mechanisms such as miscut, disorder, strain and the
non-parabolicity of the valence band,\cite{Eng07, Robert09,
Ohkawa75} but the exact sources are yet to be ascertained. Hole
density $p_{2d}$ is determined by the magnetic field $B$ at the
filling factor $\nu=5$ minimum of the longitudinal resistance curve
at T $\sim$ 1 K: $p_{2d}=\nu eB/h$, where $e$ is the electron charge
and $h$ is Planck's constant. The data (triangles) are shown in
Fig.\ \ref{fig:mobdensity}, and the dashed line is a linear fit.
Hole densities up to $7.5\times10^{11}$ cm$^{-2}$ ($V_g=-110$ V) are
obtained. If we use a parallel plate capacitor model, the equivalent
depth of the cavity is 759 nm, which is consistent with the
measurement result from a profilometer. The extrapolated threshold
voltage is $-3.9$ V, which indicates the density of the trapped
charge at the surface $\sim 2.8\times10^{10}$ cm$^{-2}$. The
mobility of the 2DHS is also shown in Fig.\ \ref{fig:mobdensity},
and the peak mobility is about 9800 cm$^2$/Vs at gate voltage
$V_g=-100$ V, T = 70 mK on this device. This is one order of
magnitude higher than the hole mobility of a Si(111)
MOSFET.\cite{KlitzSi11174} The contact resistances between the
contacts and the 2DHS increase dramatically at $|V_g|<60$ V, when
the temperature is lowered from 4.2 K to 70 mK. For example, they
increase from $\sim$ 20 K$\Omega$ to $\sim$ 400 K$\Omega$ at $V_g =
-50$ V. It is difficult to get reliable measurements at this region,
and the calculated mobility shows spurious wild fluctuations at T =
70 mK. Our discussion below will focus on the data at $|V_g|>60$ V.

\begin{figure}
\includegraphics{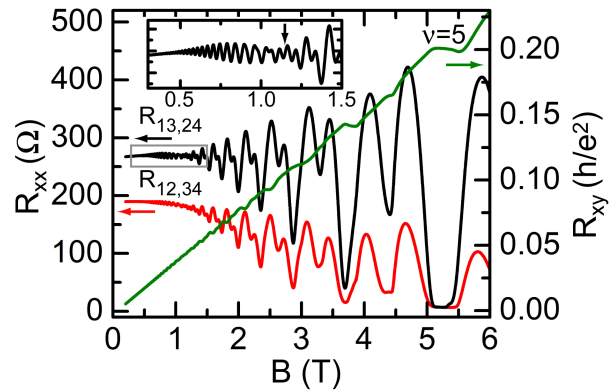}
\caption{\label{fig:magresistance} (Color online) Longitudinal
resistances $R_{13,24}$, $R_{12,34}$, and Hall resistance $R_{xy}$
vs perpendicular magnetic field $B$ at $V_g=-90$ V
($p_{2d}=6.27\times 10^{11}$ cm$^{-2}$) and T = 70 mK. Quantum Hall
effect is observed near $B=5.2$ T. The insert shows an enlarged
section of $R_{13,24}$, which exhibits a beating pattern, and the
arrow marks the node location ($B=1.14$ T).}
\end{figure}

As evidence of the high mobility 2DHS, Fig.\ \ref{fig:magresistance}
shows the longitudinal ($R_{xx}$) and Hall ($R_{xy}$) resistances as
a function of perpendicular magnetic field $B$ at $V_g=-90$ V
($p_{2d}=6.27\times 10^{11}$ cm$^{-2}$) and T = 70 mK. The integer
quantum Hall effect (IQHE) is observed. At the vicinity of $B=5.2$
T, $R_{xx}$ approaches zero, and $R_{xy}$ develops a plateau at
$\nu=5$. Clear SdH oscillations appear at $B$ field above 0.4 T. A
close look at the trace of $R_{xx}$ reveals a beating pattern of the
SdH oscillations with a node near $B=1.2$ T, as shown in the insert
of Fig.\ \ref{fig:magresistance}. Although the beating pattern has
been observed in p-channel Si(110) devices,\cite{KlitzSi11074,
Dorozhkin87} to our best knowledge, this is the first time it has
been observed on a Si(111) surface. The beating pattern can be
explained by the spin-orbit effects, although different explanations
exist. It could be from gross inhomogeneities of the carrier density
in the sample.\cite{Booth68} Two samples from different wafers have
been measured, and both show a similar beating pattern. At $V_g=-90$
V, the beating node is located at $B=1.14$ T for the device reported
here, and it is at $B=1.17$ T for the other device. It is therefore
highly unlikely that the gross inhomogeneities are the cause.

\begin{figure}
\includegraphics{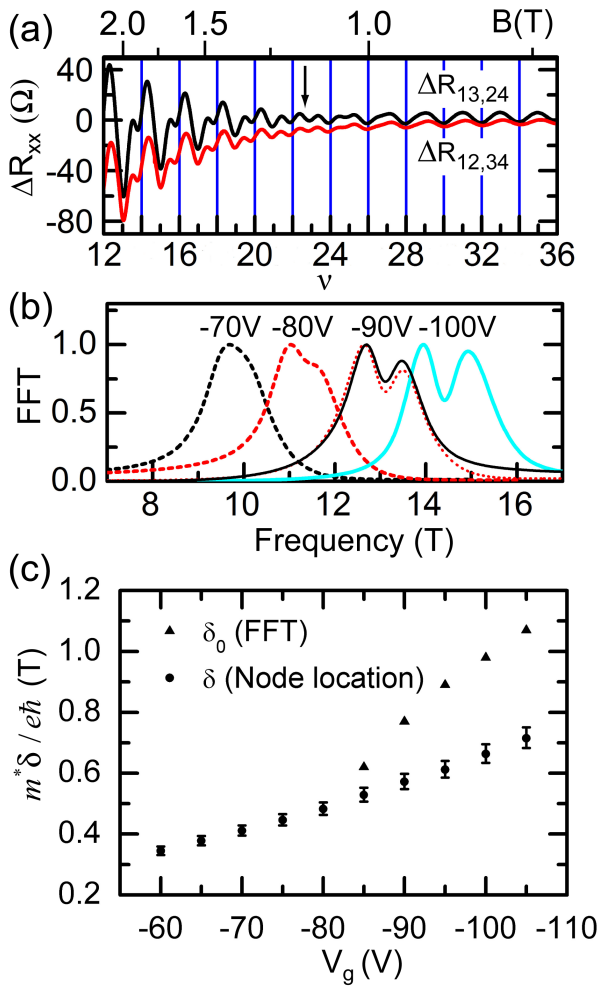}
\caption{\label{fig:spinsplitting} (Color online) (a) $\Delta
R_{13,24}$($=R_{13,24}(B)- R_{13,24}(0)$), $\Delta
R_{12,34}$($=R_{12,34}(B)- R_{12,34}(0)$) vs filling factor $\nu$ at
$V_g=-90$ V and T = 70 mK. The corresponding magnetic field $B$ is
shown on the top axis. The node location ($B=1.14$ T) is marked by
an arrow. Across the node, the minima change from the odd numbers of
$\nu$ to the even numbers of $\nu$. (b) Fourier power spectra of the
SdH oscillations ($R_{13,24}$) at different gate voltages. Maximum
peak amplitudes are normalized to 1. At $V_g=-90$ V, the solid line
is the power spectrum of $R_{13,24}$, and the dotted line is the
power spectrum of $R_{12,34}$. (c) Zero-field spin splitting and
total spin splitting (both divided by $e\hbar/m^*$) vs gate voltage.
Zero-field spin splitting is determined by Fourier power spectrum
(b). The total spin splitting is determined by the node location
(a), and the error bars are estimated by the error bound
$[\nu_{node}-1, \nu_{node}+1]$, where $\nu_{node}$ is the node
location in terms of filling factor.}
\end{figure}

The beating pattern indicates that there are two closely spaced
frequency components with similar amplitudes, arising from two
populations of holes. The hole subband densities $p^+_{2d}$ and
$p^-_{2d}$ can be determined by the Fourier analysis of the SdH
oscillations with $p^\pm_{2d}=g_sef^\pm/h$, where $g_s$ is the
degeneracy factor, $f^+$ and $f^-$ are the frequencies of the SdH
oscillations (Fig.\ \ref{fig:spinsplitting}(b)).\cite{Lu98,
WinklerSdH00} Here $R_{xx}$ vs $B^{-1}$ data in the range $0.2<B<4$
T are Fourier transformed after subtraction of a slowly varying
background, and peak locations are used as the frequencies of the
SdH oscillations. For $|V_g|\geq 85$ V (with two well-resolved
peaks), $g_s$ must be 1 in order to have the sum of the subband
densities equal to the total density, so these two components can
only be from one subband with two spin-split levels, i.e. two
spin-split subbands.

Landau levels of each spin population give rise to magnetoresistance
oscillations with similar amplitude, described by
$\cos[2\pi(E_f\pm\delta/2)/\hbar \omega_c]$, where $E_f$ is the
Fermi energy, $\delta$ is the energy separation between the two
spin-split subbands, $\hbar$ is $h/2\pi$, and $\omega_c$ is the
cyclotron frequency.\cite{Zawadzki04} The sum of these two
components leads to the modulation of the SdH oscillations given by
$A\sim \cos(\pi\delta/\hbar\omega_c)$. Beating nodes are located at
$\delta/\hbar\omega_c=n=\pm 0.5,\ \pm 1.5$, etc. Only one node is
clearly identified near $B\sim 1.2$ T (Fig.\
\ref{fig:magresistance}). Fig.\ \ref{fig:spinsplitting}(a) shows the
SdH oscillations in terms of filling factor $\nu$ at $V_g=-90$ V. At
$\nu<22.7$, the minima of the SdH oscillations are at the odd
numbers of $\nu$. At $\nu>22.7$, the minima are at the even numbers
of $\nu$. This manifests the transition across the node. We assume
that the node corresponds to $n=0.5$ (highest-field), because the
$n=0.5$ node is most easily observed, and there is no other node up
to 6 T.\cite{Datta90} The $n=1.5$ node is estimated to be around
$B\sim 0.4$ T, one third of the $n=0.5$ node, so it is difficult to
observe for this device. From the node location $B_{node}$, the
total spin splitting can be determined by $\delta=0.5\hbar
\omega_c=(e\hbar/m^*)(0.5B_{node})$, where $m^*$ is the effective
hole mass. In Fig.\ \ref{fig:spinsplitting}(c), the total spin
splitting $\delta$ is described by $m^*\delta/e\hbar=0.5B_{node}$.

In theories of the spin-orbit interaction, spin splitting can have
complicated B dependence.\cite{WinklerSO} The total spin splitting
$\delta$ can be expanded as $\delta=\delta_0+\delta_1(\hbar
\omega_c)+\delta_2(\hbar \omega_c)^2+...$, where $\delta_0$ is the
zero-field spin splitting, $\delta_1(\hbar \omega_c)$ is the linear
splitting, etc.\cite{Zawadzki04} For low magnetic fields, if only
the first two terms are considered, then $A\sim
\cos(\pi\delta_0/\hbar\omega_c+\pi\delta_1)$. The frequencies
$f^\pm$ of the SdH oscillations can be used to determined the
zero-field spin splitting by $\delta_0=(e\hbar/m^*)(f^+-f^-)$. So
the zero-field spin splitting $\delta_0$ can be described by
$m^*\delta_0/e\hbar=f^+-f^-$, also shown in Fig.\
\ref{fig:spinsplitting}(c). There are two important features in the
data. First, the zero-field spin splitting increases with increasing
perpendicular electric field $E_z$. Second, at $|V_g|\geq85$ V,
$\delta_0>\delta$, which means that a perpendicular magnetic field
can \emph{reduce} the spin splitting. More detailed analysis is
needed to better appreciate the results. These topics will be
explored in future studies.

In conclusion, the vacuum FET technique has been demonstrated to
create high mobility 2DHSs in Si(111). In these high mobility 2DHSs,
we are able to observe the IQHE, and the beating pattern of the SdH
oscillations on Si(111) surfaces, which are difficult to reach
previously. There are many interesting questions yet to be resolved,
such as the non-parabolic valence band structure and the spin-orbit
effects. Our high mobility 2DHSs open up new opportunities to
explore fundamental physics in the two-dimensional hole systems. In
addition, with the ability to create both a 2DES and a 2DHS on a
Si(111) surface, it is straightforward to develop a bipolar surface
device.

This work was funded by the Laboratory for Physical Sciences. The
authors are grateful to Oney Soykal, Charles Tahan, Kevin Eng, and
Kei Takashina for insightful discussions and technical help.


%
%

%



%

\end{document}